\documentclass[journal]{IEEEtran}
\usepackage{amsmath,amsfonts}
\usepackage{algorithmic}
\usepackage{algorithm}
\usepackage{array}

\usepackage{textcomp}
\usepackage{stfloats}
\usepackage{url}
\usepackage{verbatim}
\usepackage{graphicx}
\usepackage{cite}

\usepackage{bm}
\usepackage{url}

\usepackage{cite,graphicx,amsmath,amssymb}
\usepackage{fancyhdr}
\usepackage{hhline}
\usepackage{graphicx,graphics}
\usepackage{array,color}
\usepackage{amsmath}
\usepackage{stfloats}
\usepackage[flushleft]{threeparttable}
\usepackage{booktabs}

\newtheorem{proposition}{Proposition}
\newtheorem{remark}{Remark}

\ifCLASSINFOpdf

\else

\fi

\begin{document}
\title{Active-IRS-Enabled Target Detection}

\author{
\IEEEauthorblockN{Xianxin~Song, Xiaoqi~Qin, Xianghao~Yu, Jie~Xu, and Derrick Wing Kwan Ng}
\thanks{Xianxin~Song and Jie~Xu are with the School of Science and Engineering (SSE), the Shenzhen Future Network of Intelligence Institute (FNii-Shenzhen), and the Guangdong Provincial Key Laboratory of Future Networks of Intelligence, The Chinese University of Hong Kong (Shenzhen), Guangdong 518172, China (e-mail: xianxinsong@link.cuhk.edu.cn, xujie@cuhk.edu.cn).}
\thanks{Xiaoqi~Qin is with the State Key Laboratory of Networking and Switching Technology, Beijing University of Posts and Telecommunications, Beijing 100876, China (e-mail: xiaoqiqin@bupt.edu.cn).}
\thanks{Xianghao~Yu is with the Department of Electrical Engineering, City University of
Hong Kong, Hong Kong (e-mail: alex.yu@cityu.edu.hk).
}
\thanks{Derrick Wing Kwan Ng is with the School of Electrical Engineering and Telecommunications, University of New South Wales, Sydney, NSW 2052, Australia (e-mail: w.k.ng@unsw.edu.au).}
\thanks{Xianghao Yu and Jie Xu are the corresponding authors.}\vspace{-22pt}
}

\maketitle

\begin{abstract}
This letter studies an active intelligent reflecting surface (IRS)-enabled non-line-of-sight (NLoS) target detection system, in which an active IRS equipped with active reflecting elements and sensors is strategically deployed to facilitate target detection in the NLoS region of the base station (BS) by processing echo signals through the BS-IRS-target-IRS link. First, we design an optimal detector based on the Neyman-Pearson (NP) theorem and derive the corresponding detection probability. Intriguingly, it is demonstrated that the optimal detector can exploit both the BS's transmit signal and the active IRS's reflection noise for more effective detection. Subsequently, we jointly optimize the transmit beamforming at the BS and the reflective beamforming at the active IRS to maximize the detection probability, subject to the maximum transmit power constraint at the BS, as well as the maximum amplification power and gain constraints at the active IRS. Finally, simulation results unveil that the proposed joint beamforming design significantly enhances the detection probability, with the active IRS outperforming its fully- and semi-passive counterparts in detection performance.
\end{abstract}
\vspace{-3pt}
\vspace{-10pt}
\IEEEpeerreviewmaketitle
\section{Introduce}
\vspace{-5pt}
Recently, intelligent reflecting surface (IRS) has emerged as a promising technology for future sixth-generation (6G) wireless networks, aiming to enhance both sensing and communication capabilities by dynamically reconfiguring the wireless transmission environment \cite{10422881,10077119,10243495}. Specifically, IRSs can establish additional reflection links from the base station (BS) transceiver to the IRS and then to the sensing target. These links can be leveraged to realize non-line-of-sight (NLoS) target sensing\cite{10138058} and multi-angle sensing\cite{9732186}, thereby effectively extending the sensing region and enhancing sensing performance in both target detection and estimation.

This letter studies a fundamental IRS-enabled sensing system, in which an IRS is deployed to establish a virtual line-of-sight (LoS) link between the BS and the sensing target, thus realizing NLoS target detection. In such IRS-enabled NLoS target detection system, designing the detector and optimizing the joint beamforming are critical, posing significant challenges due to multiple signal reflections and the intricacies of the coupled variables in the associated non-convex optimization design problem. There are generally three different types of IRSs applicable for sensing, namely fully-passive IRS, semi-passive IRS, and active IRS. The fully-passive IRS consists of passive reflecting elements that can alter the phase shift of incident signals without introducing any amplification. As a result, the corresponding target sensing is implemented at the BS by processing echo signals through the BS-IRS-target-IRS-BS link \cite{10138058,9732186}. Next, the semi-passive IRS, equipped with dedicated sensors for signal processing, conducts target sensing at the IRS, exploiting echo signals through the BS-IRS-target-IRS link \cite{9724202,xianxin}. Furthermore, the active IRS is deployed with active reflecting elements for adaptative signal amplification and phase adjustment along with dedicated sensors. Indeed, the BS-IRS-target reflection link in practice often suffers from distance-product path loss, which significantly restricts sensing performance. Consequently, the active IRS presents great potential to efficiently compensate for this substantive path loss by amplifying the incident signal. In the literature, various prior works \cite{10138058,9732186,9724202,xianxin,fang2024joint,10186271,9979782,10054402,10319318,10496515} have investigated fully-passive IRS sensing \cite{10138058,9732186}, semi-passive IRS sensing \cite{9724202,xianxin}, and active IRS sensing\cite{fang2024joint} and ISAC\cite{10186271,9979782,10054402,10319318,10496515}, with a handful of works studying at active-IRS-enabled wireless communications\cite{9998527,9377648}. However, the above prior works about active-IRS-enable sensing and ISAC adopted target illumination power \cite{10186271} and beampattern match error \cite{9979782} as sensing performance metrics, or treated the active IRS’s reflection noise as harmful interference\cite{10054402,fang2024joint,10319318,10496515}. Indeed, the potential roles of active IRS's reflection noise in detection, the optimal detector that exploits both the BS's transmit signal and the IRS's reflection noise, and the design of joint beamforming are areas that remain unexplored. These research gaps serve as the motivation for our investigation in this work.

In this letter, we consider target detection enabled by an active IRS, in which the active IRS is deployed to detect the  presence of a target by processing echo signals from the BS-IRS-target-IRS link. First, we design an optimal detector based on the Neyman-Pearson (NP) theorem. It is demonstrated that the obtained optimal detector jointly harnesses both the BS's transmit signal and the active IRS's reflection noise for effective target detection. Then, we derive the detection probability under a predetermined false alarm probability. Next, we jointly optimize the transmit beamforming at the BS, alongside the phase shift and amplification matrices at the active IRS, to maximize the derived detection probability. This optimization adheres to the maximum transmit power constraint at the BS and the maximum amplification power and gain constraints at the active IRS. Finally, simulation results demonstrate that the proposed joint beamforming design achieves higher detection probability, verifying that the adoption of an active IRS can significantly increase the  detection performance compared with fully- and semi-passive counterparts.

\textit{Notations:} 
Boldface letters refer to vectors (lower case) or matrices (upper case). For a square matrix $\mathbf S$, $\mathrm {tr}(\mathbf S)$ denotes its trace and $\mathbf S \succeq \mathbf{0}$ means that $\mathbf S$ is a positive semi-definite matrix. For an arbitrary-size matrix $\mathbf M$, $\mathbf M^*$, $\mathbf M^{T}$, and $\mathbf M^{H}$ denote its conjugate, transpose, and conjugate transpose, respectively. Let $\mathcal{CN}(\mathbf 0,\mathbf \Sigma)$ denote the distribution of a circularly symmetric complex Gaussian (CSCG) random vector with mean vector $\mathbf 0$ and covariance matrix $\mathbf \Sigma$, and $\sim$ denote “distributed as”. The spaces of $x \times y$ real and complex matrices are denoted by $\mathbb{R}^{x \times y}$ and $\mathbb{C}^{x \times y}$, respectively. Let $\mathrm {diag}(a_1,\cdots,a_N)$ denote a diagonal matrix with diagonal elements $a_1,\cdots,a_N$, $\otimes$ denote the Kronecker product, $\|\cdot\|$ denote the Euclidean norm, and $\mathbb{E}(\cdot)$ denote the statistical expectation.

\vspace{-15pt}
\section{System Model}
\begin{figure}[t]
    \centering
        \includegraphics[width=0.25\textwidth]{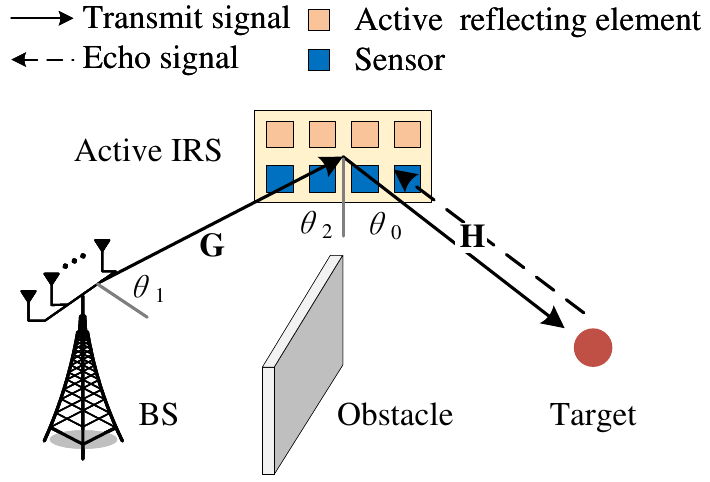}
      \caption{Illustration of active IRS-enabled target detection.}
      \label{system_model}\vspace{-15pt}
\end{figure}

\vspace{-5pt}
As shown in Fig.~\ref{system_model}, we consider an active IRS-enabled target detection system consisting of one BS with $M_t$ uniform linear array (ULA) transmit antennas, one active IRS with $N$ ULA active reflecting elements and $M_r$ ULA sensors, and one potential target located at the BS's NLoS region\footnote{This work can be extended to the multi-target case by extending the two hypotheses tests to multiple hypotheses tests.}. We consider a binary detection task, which aims to decide the presence of a target by processing echo signals from the BS-IRS-target-IRS link. Let hypotheses $\mathcal{H}_1$ and $\mathcal{H}_0$ denote the cases when the target is present and absent, respectively.

First, we introduce the channel models. We assume that the BS-IRS and IRS-target-IRS links are dominated by the LoS links \cite{10155421,10178078}. Without loss of generality, we assume that the distance between adjacent transmit antennas at the BS, as well as that between adjacent reflecting elements or sensors at the IRS are all equal to half carrier wavelength. Let vector $\mathbf e(\varsigma, K)\in\mathbb C^{K\times 1}$ denote the steering vector function of a ULA with size $K$ towards any angle $\varsigma$, i.e., $\mathbf e(\varsigma, K) = [1, e^{j\pi\sin \varsigma},\cdots,e^{j\pi(K-1)\sin \varsigma}]^T$. The angle of the IRS with respect to (w.r.t.) the BS's transmit antennas and the angle of the BS w.r.t. the IRS's reflecting elements are denoted as $\theta_1$ and $\theta_2$, respectively. The distances of the BS-IRS and IRS-target links are denoted as $d_1$ and $d_2$, respectively. As such, the channel matrix of the BS-IRS link is given as $\mathbf G = \sqrt{L(d_1)} \mathbf e(\theta_2,N)\mathbf e^T(\theta_1,M_t)\in \mathbb{C}^{N\times M_t}$, where $L(d_1)$ is the associated channel path loss. Similarly, the target response matrix of the IRS-target-IRS link is modeled as $\mathbf H = \alpha\mathbf e(\theta_0,M_r) \mathbf e^T(\theta_0,N)\in \mathbb{C}^{M_r\times N}$, where $\alpha= \beta L(d_2) \in \mathbb C$ is the channel coefficient depending on both the radar cross section (RCS) of the target $\beta$ and the path loss of the IRS-target link $L(d_2)$\cite{10138058,10496515}, and $\theta_0$ is the angle of target w.r.t. the IRS\footnote{In this work, we focus on the target detection at a specified location, based on which the target coefficient $\alpha$ and angle $\theta_0$ can be predicted {\it a priori}. As such, we assume that $\alpha$ and $\theta_0$ are known to facilitate the analysis.}.

Next, we present the signal transmission model. We focus on the target detection over a block with a total of $T$ sensing symbols. Let $\mathcal{T} = \{1,\cdots,T\}$ denote the set of symbols in the block. The transmit signal by the BS at symbol $t \in \mathcal{T}$ is denoted as $\mathbf x(t) \in \mathbb{C}^{M_t\times 1}$ with sample covariance matrix $\mathbf R_x = \frac{1}{T} \sum_{t=1}^{T}\mathbf x(t) \mathbf x^H(t)\succeq \mathbf 0$\cite{10138058,9724202}. Note that the transmit signal $\mathbf x(t)$ is deterministic and thus known by the IRS sensors prior to target sensing. Let $P$ denote the maximum BS transmit power, i.e., $\frac{1}{T} \sum_{t=1}^{T}\|\mathbf x(t)\|^2=\mathrm{tr}(\mathbf R_x)\le P$. Also, the phase shift matrix at the active IRS is denoted as $\mathbf \Phi=\mathrm{diag}(e^{j\phi_1},\cdots,e^{j\phi_N})\in \mathbb{C}^{N\times N}$ with $\phi_n\in(0,2\pi], \forall n\in \mathcal{N}$, and the amplification matrix at the active IRS is denoted as $\mathbf A=\mathrm{diag}(a_1,\cdots,a_N)\in \mathbb{R}^{N\times N}$ with $0 \le a_n\le a_{\max}, \forall n\in \mathcal{N}$, where $a_{\max}$ is the maximum amplification gain at each active reflecting element. After the signal transmission from the BS to the IRS and the IRS's reflection, the departed signal from the active IRS is $\mathbf A\mathbf \Phi\mathbf G \mathbf x(t) + \mathbf A\mathbf \Phi\mathbf z(t)$, where $\mathbf z(t) \sim \mathcal{CN}(\mathbf 0,\sigma_z^2\mathbf I_{N})$ is the active IRS’s reflection noise and $\sigma_z^2$ is the inherent noise power at each IRS's reflecting element. Let $P_A$ denote the maximum IRS amplification power,  i.e., $\frac{1}{T} \sum_{t=1}^{T}\|\mathbf A\mathbf \Phi\mathbf G \mathbf x(t) + \mathbf A\mathbf \Phi\mathbf z(t)\|^2=(L(d_1)\mathbf e^T(\theta_1,M_t) \mathbf R_x\mathbf e^*(\theta_1,M_t) + \sigma_z^2)\sum_{n=1}^Na_n^2\le P_A$. 
Accordingly, the received echo signals by the active IRS under hypotheses $\mathcal{H}_0$ and $\mathcal{H}_1$ at symbol $t\in\mathcal{T}$ are respectively given by 
\begin{equation}
\begin{cases}
\mathcal{H}_0:&\mathbf y (t) = \mathbf n(t),\\
\mathcal{H}_1:&\mathbf y (t) = \alpha\mathbf e (\theta_0,M_r) \mathbf e^T(\theta_0,N)\mathbf A\mathbf \Phi\mathbf G \mathbf x(t)\\
&\qquad + \alpha\mathbf e(\theta_0,M_r)\mathbf e^T(\theta_0,N)\mathbf A\mathbf \Phi\mathbf z(t) + \mathbf n(t),
\end{cases}
\end{equation}
where $\mathbf n(t) \in \mathbb{C}^{M_r\times 1}\sim\mathcal{CN}(\mathbf 0,\sigma^2\mathbf I_{M_r})$ is the noise at the IRS sensors with $\sigma^2$ being the noise power at each IRS's sensor.
Furthermore, we stack the received each signal by the active IRS, the transmitted signal by the BS, the active IRS's reflection noise, and noise at the IRS's receiver over the $T$ symbols as 
$\tilde{\mathbf y} = [\mathbf y^T(1), \cdots,\mathbf y^T(T)]^T$,
$\tilde{\mathbf x} = [\mathbf x^T(1), \cdots,\mathbf x^T(T)]^T$,
$\tilde{\mathbf z} = [\mathbf z^T(1), \cdots,\mathbf z^T(T)]^T$, and
$\tilde{\mathbf n} = [\mathbf n^T(1), \cdots,\mathbf n^T(T)]^T$, respectively.
Then, the received echo signals at the active IRS under hypotheses $\mathcal{H}_0$ and $\mathcal{H}_1$ over the $T$ symbols are respectively given by
\begin{equation}\label{eq:echo_signal_t}
\begin{cases}
\mathcal{H}_0:&\tilde{\mathbf y} = \tilde{\mathbf n},\\
\mathcal{H}_1:&\tilde{\mathbf y} =(\mathbf I_{T}\otimes \alpha\mathbf e (\theta_0,M_r) \mathbf e^T(\theta_0,N)\mathbf A\mathbf \Phi\mathbf G) \tilde{\mathbf x}\\
&\qquad + (\mathbf I_{T}\otimes \alpha\mathbf e(\theta_0,M_r)\mathbf e^T(\theta_0,N)\mathbf A\mathbf \Phi)\tilde{\mathbf z} + \tilde{\mathbf n}.
\end{cases}
\end{equation}

Based on the received echo signals in \eqref{eq:echo_signal_t}, we aim to detect the presence of the target. The difference of received signals between hypotheses $\mathcal{H}_0$ and $\mathcal{H}_1$ consists the deterministic component signal $\mathbf u_1 = (\mathbf I_{T}\otimes \alpha\mathbf e (\theta_0,M_r) \mathbf e^T(\theta_0,N)\mathbf A\mathbf \Phi\mathbf G) \tilde{\mathbf x}$, which is induced by the BS's transmit signal, and random component signal $\mathbf u_2 = (\mathbf I_{T}\otimes \alpha\mathbf e(\theta_0,M_r)\mathbf e^T(\theta_0,N)\mathbf A\mathbf \Phi)\tilde{\mathbf z}$, which is induced by the IRS's reflection noise. It is thus essential to jointly leverage both deterministic and random signals to fully exploit the available signal dimensions for effective sensing, highlighting the pressing need to optimize transmit and reflective beamforming for enhancing the sensing performance.

\section{Optimal Target Detector}\label{sec:detector_design_and_detection_probability}
In this section, we first obtain the optimal detector based on \eqref{eq:echo_signal_t} via the NP theorem\cite{steven1993fundamentals} and then derive the target detection probability under given false alarm probability. 

We denote the covariance matrix of the random signal $\mathbf u_2$ as $\mathbf C = \mathbb{E}(\mathbf u_2 \mathbf u_2^H)  =\mathbf I_{T}\otimes \sigma_z^2|\alpha|^2\sum_{n=1}^N a_n^2\mathbf e(\theta_0,M_r)\mathbf e^H(\theta_0,M_r)$. Then, the probability density functions (PDFs) of the received signal $\tilde{\mathbf y}$ under hypotheses $\mathcal{H}_0$ and $\mathcal{H}_1$ are given as
$p(\tilde{\mathbf y};\mathcal{H}_0) = \frac{\exp\left(-\frac{1}{\sigma^2}\tilde{\mathbf y}^H\tilde{\mathbf y}\right)}{\pi^{M_tT}\det(\sigma^2\mathbf I_{M_rT})}$ and $p(\tilde{\mathbf y};\mathcal{H}_1) = \frac{\exp\left(-(\tilde{\mathbf y} - \mathbf u_1)^H(\mathbf C+\sigma^2\mathbf I_{M_rT})^{-1}(\tilde{\mathbf y} - \mathbf u_1)\right)}{\pi^{M_tT}\det(\mathbf C+\sigma^2\mathbf I_{M_rT})}$,  respectively.
For the binary detection task, we adopt the NP theorem to establish the optimal detection rule\cite{steven1993fundamentals}, which aims to determine the optimal decision threshold to maximize the detection probability under a predetermined false alarm probability. The NP detector is $\frac{p(\tilde{\mathbf y};\mathcal{H}_1)}{p(\tilde{\mathbf y};\mathcal{H}_0)}\stackrel{\mathcal{H}_1}{\underset{\mathcal{H}_0}{\gtrless}} \delta$,
which is equivalent to $\ln\frac{p(\tilde{\mathbf y};\mathcal{H}_1)}{p(\tilde{\mathbf y};\mathcal{H}_0)}
=\tilde{\mathbf y}^H\frac{1}{\sigma^2}\mathbf C(\mathbf C+\sigma^2\mathbf I_{M_rT})^{-1}\tilde{\mathbf y}
- \mathbf u_1^H(\mathbf C+\sigma^2\mathbf I_{M_rT})^{-1}\mathbf u_1
+2\mathrm{Re}\left\{\mathbf u_1^H(\mathbf C+\sigma^2\mathbf I_{M_rT})^{-1}\tilde{\mathbf y}\right\}
+ M_rT\ln \sigma^{2}-\ln \det(\mathbf C+\sigma^2\mathbf I_{M_rT})\stackrel{\mathcal{H}_1}{\underset{\mathcal{H}_0}{\gtrless}} \ln\delta$,
where the detection threshold $\delta$ is determined with the given false alarm probability $P_\text{FA}$\footnote{The closed-form relation between $\delta$ and $P_\text{FA}$ is derived in Appendix~\ref{sec:Proof_of_Lemma_1}.}. By letting $\delta' = \ln\delta + \mathbf u_1^H(\mathbf C+\sigma^2\mathbf I_{M_rT})^{-1}\mathbf u_1-M_rT\ln \sigma^{2}+\ln \det(\mathbf C+\sigma^2\mathbf I_{M_rT})$, we obtain the optimal detector as 
$T(\tilde{\mathbf y}) 
=\tilde{\mathbf y}^H\frac{1}{\sigma^2}\mathbf C(\mathbf C+\sigma^2\mathbf I_{M_rT})^{-1}\tilde{\mathbf y}
+2\mathrm{Re}\left\{\mathbf u_1^H(\mathbf C+\sigma^2\mathbf I_{M_rT})^{-1}\tilde{\mathbf y}\right\}\stackrel{\mathcal{H}_1}{\underset{\mathcal{H}_0}{\gtrless}} \delta'$.
\begin{remark}
For the optimal detector $T(\tilde{\mathbf y})$, both the BS's transmit signal $\mathbf x(t)$ and IRS’s reflection noise $\mathbf z(t)$ are effectively utilized for detection. Therefore, it is expected that the IRS’s reflection noise can be harnessed to enhance the detection performance, as will be shown in Section~\ref{sec:numerical_results}.
\end{remark}  

 Then, we have the following proposition to analyze the detection performance.
\begin{proposition}\label{lm:P_D_under_P_F}
The detection probability $P_\text{D}(\mathbf R_x, \mathbf \Phi, \mathbf A)$  under a predetermined false alarm probability $P_\text{FA}$ is 
\begin{equation}\label{eq:P_D_given_P_F}
P_\text{D}(\mathbf R_x, \mathbf \Phi, \mathbf A) = \mathcal{Q}_{\chi^2_{2T}(\lambda_2)}\left(\frac{\mathcal{Q}^{-1}_{\chi^2_{2T}(\lambda_1)}\left(P_\text{FA}\right)}{1+|\alpha|^2M_r\!\sum_{n=1}^N a_n^2\sigma_z^2/\sigma^2}\right),
\end{equation}
with
$\lambda_1=\frac{2\sigma^2T\mathbf e^T(\theta_0,N)\mathbf A\mathbf \Phi\mathbf G \mathbf R_x \mathbf G^H \mathbf \Phi^H\mathbf A^H \mathbf e^*(\theta_0,N)}{(\sum_{n=1}^Na_n^2)^2M_r\sigma_d^4|\alpha|^2}$ and $\lambda_2=\lambda_1\big(1\!+\!|\alpha|^2\sum_{n=1}^Na_n^2M_r\sigma_z^2/\sigma^2\big)$ being the non-centrality parameters of non-central chi-squared distributions.
\end{proposition}
\begin{IEEEproof}
Please refer to Appendix~\ref{sec:Proof_of_Lemma_1}.
\end{IEEEproof}

Proposition~\ref{lm:P_D_under_P_F} reveals the achieved detection probability, which, however, is a highly non-convex and non-explicit function due to the right-tail function of non-central chi-squared distributions in \eqref{eq:P_D_given_P_F}. As such, it is difficult to adopt $P_\text{D}\left(\mathbf R_x, \mathbf \Phi, \mathbf A \right)$ as the objective function directly for optimizing the joint beamforming. To tackle this issue, we further approximate the detection probability by considering the special asymptotical case when the number of symbols $T$ is sufficiently large. 
\begin{proposition}\label{le:P_D_approximation}
When $T$ becomes sufficiently large, the detection probability in \eqref{eq:P_D_given_P_F} approaches \eqref{eq:P_D_approximation} as given at the top of next page.
\begin{IEEEproof}
According to the central limit theorem, the PDF of chi-squared distribution becomes Gaussian when the degrees of freedom, $2T$, are sufficiently large\cite{steven1993fundamentals}. Thus, the detection probability in \eqref{eq:P_D_given_P_F} is approximated as \eqref{eq:P_D_approximation}. The detailed proof is given in Appendix~\ref{Sec:proof_lemma_2}.
\end{IEEEproof}
\end{proposition}

\begin{figure*}[t]
\begin{small}
\begin{equation}\label{eq:P_D_approximation}
\tilde{P}_\text{D}\left(\mathbf R_x, \mathbf \Phi, \mathbf A\right) \!=\! \mathcal{Q}\left(\frac{\sqrt{T+\lambda_1}\mathcal{Q}^{-1}\left(\tilde{P}_\text{FA}\right)\!-\!|\alpha|^2M_r\sum_{n=1}^Na_n^2T\sigma_z^2/\sigma^2\!-\!\lambda_1\left((1+|\alpha|^2M_r\sum_{n=1}^Na_n^2\sigma_z^2/\sigma^2)^2-1\right)/2}{\left(1+|\alpha|^2M_r\sum_{n=1}^Na_n^2\sigma_z^2/\sigma^2\right)\sqrt{T+\lambda_1\left(1+|\alpha|^2M_r\sum_{n=1}^Na_n^2\sigma_z^2/\sigma^2\right)}}\right).
\end{equation}
\end{small}
\hrulefill
\vspace{-18pt}
\end{figure*}
We have the following interesting observation based on Proposition~\ref{le:P_D_approximation}. 
\begin{proposition}\label{le:P_D_lambda}
Given fixed IRS amplification matrix $\mathbf A$, the  detection probability $\tilde{P}_\text{D}\left(\mathbf R_x, \mathbf \Phi, \mathbf A\right)$ is  monotonically increasing  w.r.t. the non-centrality parameter $\lambda_1$.
\end{proposition}
\begin{IEEEproof}
This lemma follows directly by calculating the derivation of $\tilde{P}_\text{D}\left(\mathbf R_x, \mathbf \Phi, \mathbf A\right)$ w.r.t. $\lambda_1$, which is larger than zero.
\end{IEEEproof}

\vspace{-5pt}
\section{Target Detection Probability Maximization via Joint Beamforming}\label{sec:Joint_BF}
\vspace{-3pt}

In this section, we aim to maximize the detection probability by jointly optimizing the transmit beamforming at the BS, as well as the phase shift and amplification matrices at the active IRS. By exploiting the approximate detection probability in \eqref{eq:P_D_approximation} as the objective function, the detection probability maximization problem is formulated as 
\begin{subequations}
  \begin{align}\notag
    &\text{(P1)}:\max_{\mathbf R_x\succeq \mathbf 0, \mathbf \Phi, \mathbf A}\quad  \tilde{P}_\text{D}\left(\mathbf R_x, \mathbf \Phi, \mathbf A\right) \\ \label{eq:unit_phi}
    \text { s.t. }& \quad  |\mathbf \Phi_{n,n}|=1, \forall n\in \mathcal{N},\\ \label{eq:a_max}
    &\quad 0 \le a_n \le a_{\max}, \forall n\in \mathcal{N},\\ \label{eq:IRS_power_constraint}
    &\quad (L(d_1)\mathbf e^T(\theta_1,M_t) \mathbf R_x\mathbf e^*(\theta_1,M_t) + \sigma_z^2)\sum_{n=1}^Na_n^2\le P_A,\\\label{eq:BS_power_constraint}
    &\quad \mathrm{tr}(\mathbf R_x) \le P.
  \end{align}
\end{subequations}

In the following, we adopt nested optimization to solve problem (P1) optimally. First, we derive optimal $\mathbf R_x$ and $\mathbf \Phi$ in a closed form under any given $\mathbf A$. Based on Proposition~\ref{le:P_D_lambda}, maximizing $\tilde{P}_\text{D}\left(\mathbf R_x, \mathbf \Phi, \mathbf A\right)$ is equivalent to maximizing $\lambda_1$ under given $\mathbf A$. Therefore,  the optimization of $\mathbf R$ and $\mathbf \Phi$ is formulated as
\begin{subequations}
  \begin{align}\notag
    \text{(P2)}:\max_{\mathbf R_x\succeq \mathbf 0, \mathbf \Phi}&\quad  \mathbf e^T(\theta_0,N)\mathbf A\mathbf \Phi\mathbf G \mathbf R_x \mathbf G^H \mathbf \Phi^H\mathbf A^H \mathbf e^*(\theta_0,N)\\ \notag
    \text { s.t. }& \quad  \eqref{eq:unit_phi},~\eqref{eq:IRS_power_constraint},~\text{and}~\eqref{eq:BS_power_constraint}.
  \end{align}
\end{subequations}
\begin{proposition}\label{pr:optimal_transmit}
The optimal solution to problem (P2) is $\mathbf R^\star_x(\mathbf A) = \frac{P_x}{M_t}\mathbf e^*(\theta_1,M_t)\mathbf e^T(\theta_1,M_t)$ with $P_x =\min\big\{P, (P_A/\sum_{n=1}^Na_n^2-\sigma_z^2)/(L(d_1)M_t)\big\}$ and $\mathbf{\Phi}^\star = \mathrm {diag}(e^{j\phi_1^\star},\cdots,e^{j\phi_{N}^\star})$ with $\phi_n^\star = -j2\pi(n-1)d_2(\sin \theta_0+\sin \theta_2)/\lambda, \forall n \in \mathcal{N}$.
\begin{IEEEproof}
The optimal transmit beamforming at the BS is obtained based on the maximum ratio transmission. Besides, the optimal reflecting phase shift design is achieved by aligning the multi-path signals toward the sensing target.
\end{IEEEproof}
\end{proposition}

Next, we substitute the obtained $\mathbf R^\star_x(\mathbf A)$ and $\mathbf{\Phi}^\star$ back to problem (P1) and then optimize $\mathbf A$, which is formulated as
\setcounter{equation}{5} 
\begin{subequations}
  \begin{align}\notag
    \text{(P3)}:\max_{\mathbf A, P_x}&\quad  \tilde{P}_\text{D}\left(\mathbf R^\star_x(\mathbf A), \mathbf \Phi^\star, \mathbf A\right)\\
    \text { s.t. }& \quad 0 \le a_n \le a_{\max}, \forall n\in \mathcal{N},\\\label{eq:power_IRS_a_1}
    & \quad  \left(L(d_1)M_t P_x + \sigma_z^2\right)\sum_{n=1}^Na_n^2\le  P_A,\\\label{eq:power_IRS_a_2}
    & \quad  0 \le P_x \le P.
  \end{align}
\end{subequations}

We have the following proposition for revealing problem (P3). 
\begin{proposition}\label{pro:optimality_A}
The optimality of problem (P3) is attained when $a_n=a_0, \forall n\in\mathcal{N}$.
\end{proposition}
\begin{IEEEproof}
With the optimal $\mathbf R^\star_x$ and $\mathbf{\Phi}^\star$, we have $\lambda_1=\frac{2\sigma^2TL(d)M_tP_x(\sum_{n=1}^N a_n)^2}{(\sum_{n=1}^Na_n^2)^2M_r\sigma_d^4|\alpha|^2}$.
%By substituting $\mathbf R^\star_x(\mathbf A)$ and $\mathbf{\Phi}^\star$ back to $\tilde{P}_\text{D}\left(\mathbf R_x, \mathbf \Phi, \mathbf A\right)$, we optimize the reflecting amplification matrix $\mathbf A$.
For variable $\lambda_1$, with any given amplification matrix $\mathbf A$, we all have $(\sum_{n=1}^N a_n)^2\le N\sum_{n=1}^Na_n^2$, where the inequality holds if and only if $a_n=a_0, \forall n\in \mathcal{N}$. Then, as the detection probability $\tilde{P}_\text{D}\left(\mathbf R_x, \mathbf \Phi, \mathbf A\right)$ increases monotonically w.r.t. $\lambda_1$, Proposition~\ref{pro:optimality_A} is obtained.
\end{IEEEproof}

As variable $\lambda_1$ is monotonically increasing  w.r.t. $P_x$, based on Proposition~\ref{le:P_D_lambda}, the optimality of problem (P3) is attained when at least one of the constraints \eqref{eq:power_IRS_a_1} and \eqref{eq:power_IRS_a_2} is tight. Then, applying Proposition~\ref{pro:optimality_A}, problem (P3) can be optimally solved by a one-dimensional search of the feasible region of variable $a_0$. Therefore, problem (P1) is optimally solved. 
\vspace{-10pt}
\section{Numerical Results}\label{sec:numerical_results}
\vspace{-5pt}
In this section, we present simulation results to verify the detection performance of our proposed detector and beamforming design. We set the numbers of transmit antennas at the BS and the receive sensors at the active IRS as $M_t =M_r =8$. The distance-dependent path loss at distance $d$ is modeled as $L(d) = K_0(d/d_0)^{-\beta}$ with $K_0 =-30$~dB being the path loss at the reference distance $d_0 =1~\text{m}$ and $\beta=2.2$. The distances of the BS-IRS and IRS-target links are set as $d_1 = 120$~m and $d_2=10$~m, respectively. The target’s RCS is set as $\beta=1$. The false alarm probability is set as $\tilde{P}_\text{FA} = 10^{-3}$. We also set $\theta_0=\theta_1 = \theta_2 = \pi/4$, $P=30$~dBm, $P_A = 15$~dBm, $\sigma^2=-70$~dBm, and $\sigma_z^2=-30$~dBm. For comparison, we also consider the following benchmark schemes, as well as the fully- and semi-passive IRS-enabled target detection \cite{xianxin} with the maximum transmit power at the BS being $P+P_A$. 

\subsubsection{Sensing signal-to-noise ratio (SNR)-based detector} In this benchmark scheme, the detector is  designed based on the received sensing SNR 
$\gamma=\frac{|\alpha|^2\mathbf e^T(\theta_0,N)\mathbf A\mathbf \Phi\mathbf G \mathbf R_x \mathbf G^H \mathbf \Phi^H\mathbf A^H \mathbf e^*(\theta_0,N)}{|\alpha|^2\mathbf e^T(\theta_0,N)\mathbf A\mathbf \Phi\mathbf \Phi^H\mathbf A^H \mathbf e^*(\theta_0,N)\sigma_z^2+\sigma^2}$.
This approach is based on  that in semi-passive IRS-enabled target detection\cite{xianxin}. Accordingly, the joint beamforming is optimized to maximize the sensing SNR $\gamma$.

\subsubsection{Reflective beamforming only} We leverage the proposed detector in Section~\ref{sec:detector_design_and_detection_probability} and only optimize the amplification and reflecting phase shift matrices at the IRS with the sample covariance matrix $\mathbf R_x = P/M_t\mathbf I_{M_t}$, to maximize the target detection probability in \eqref{eq:P_D_approximation}.

\subsubsection{Transmit beamforming only} We adopt the proposed detector in Section~\ref{sec:detector_design_and_detection_probability} and only optimize the transmit beamforming at the BS with the maximum IRS amplification matrix $\mathbf A =\mathrm{diag}(a'_{\max},\cdots,a'_{\max})$, $a'_{\max}=\min(a_{\max},\sqrt{P_A/(N\sigma_z^2)})$ and adopt a random phase shift matrix, to maximize the target detection probability in \eqref{eq:P_D_approximation}. 

% \begin{figure}[t]
%     \centering
%         \includegraphics[width=0.4\textwidth]{P_d_T_final_4.pdf}
%       \caption{The detection probability versus the number of sensing symbols $T$, where $N =16$.}
%       \label{P_d_distance}
% \end{figure}
% \begin{figure}[t]
%     \centering
%         \includegraphics[width=0.4\textwidth]{P_d_elements_final_4.pdf}
%       \caption{The detection probability versus the number of reflecting elements $N$, where $T =8$.}
%       \label{P_d_elements}\vspace{-15pt}
% \end{figure}

\begin{figure*}
\centering
\begin{minipage}{0.6\columnwidth}\centering
\centering\includegraphics[scale=0.4]{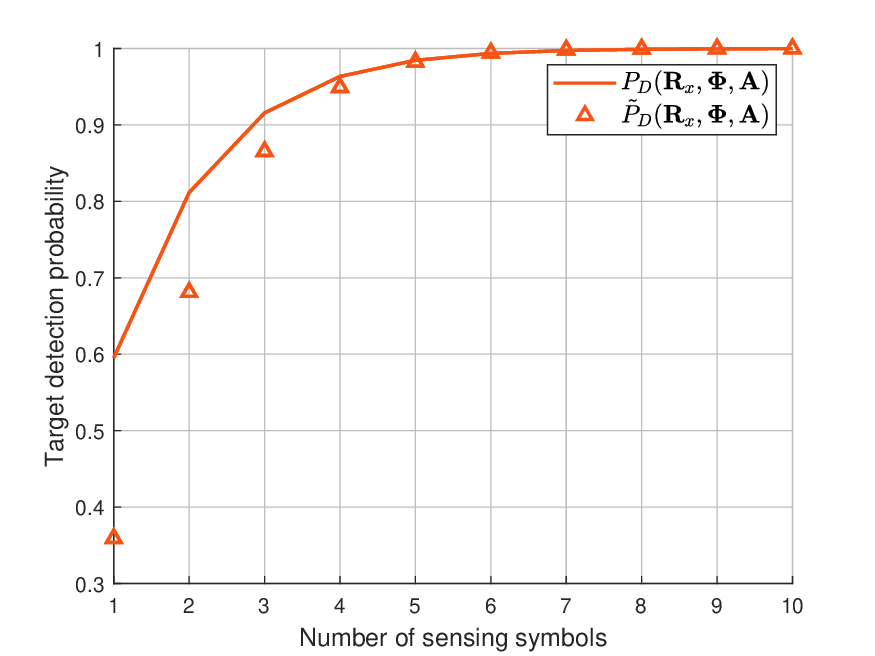}\vspace{-10pt}\caption{\label{P_d_approximation}The detection probability versus the number of sensing symbols $T$, where $N =16$ and $a_{\max} = 20$~dB.}
\end{minipage}\hspace{+0.6cm}
\begin{minipage}{0.6\columnwidth}\centering
\centering\includegraphics[scale=0.4]{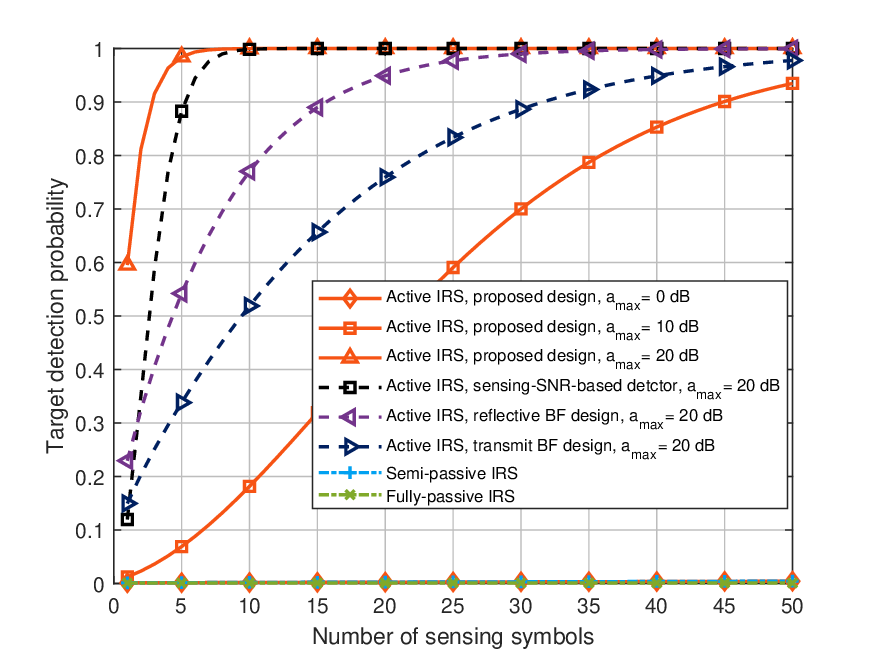}\vspace{-10pt}\caption{\label{P_d_distance}The detection probability versus the number of sensing symbols $T$, where $N =16$.}
\end{minipage}\hspace{+0.6cm}
\begin{minipage}{0.6\columnwidth}\centering
\centering\includegraphics[scale=0.4]{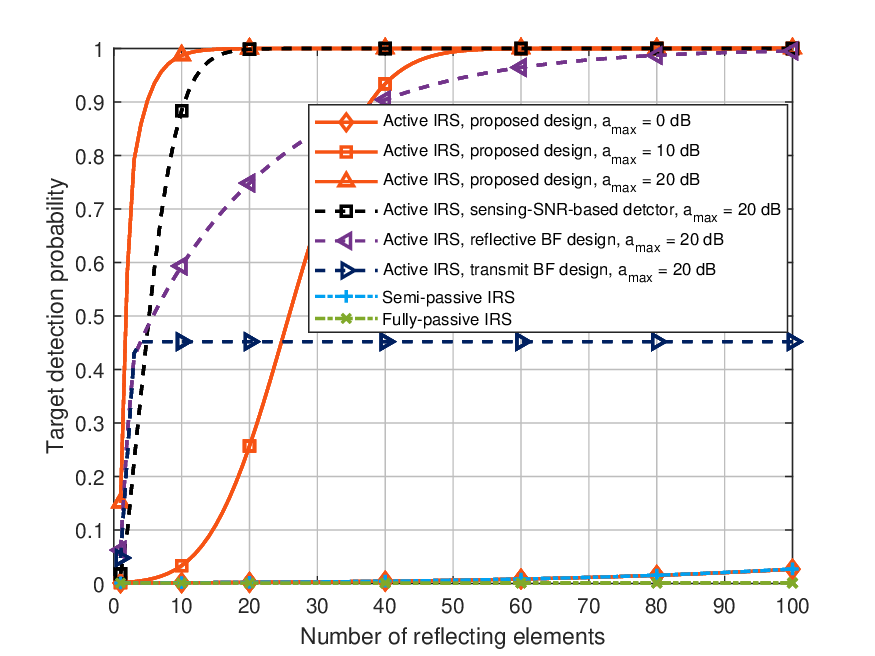}\vspace{-10pt}\caption{\label{P_d_elements}The detection probability versus the number of reflecting elements $N$, where $T =8$.}
\end{minipage}\vspace{-0.4cm}
\end{figure*}

Figs.~\ref{P_d_approximation}, \ref{P_d_distance} and \ref{P_d_elements} illustrate the detection probabilities versus the number of sensing symbols $T$ and the number of reflecting elements $N$, respectively. It is demonstrated that the detection probability $P_\text{D}(\mathbf R_x, \mathbf \Phi, \mathbf A)$ in \eqref{eq:P_D_given_P_F} closely approximates as $\tilde{P}_\text{D}\left(\mathbf R_x, \mathbf \Phi, \mathbf A\right)$ in \eqref{eq:P_D_approximation} when the the number of sensing symbols $T\ge 5$, thus verifying the accuracy of our approximation. It is also shown that our proposed joint beamforming design outperforms the benchmark adopting the sensing-SNR-based detector. This superiority stems from our design's ability to exploit both the BS's transmit signal and the active IRS's reflection noise, while the sensing-SNR-based detector only utilizes the BS's transmit signal and treats the active IRS’s reflection noise as harmful interference. Meanwhile, it is also observed that our proposed design with active IRS achieves significantly higher detection probability than the counterparts with semi-passive and fully-passive IRSs. The results show the advantages of signal amplification at the active  IRS. Furthermore, it is observed that our proposed joint beamforming design achieves enhanced sensing performance as compared to the benchmark schemes with only reflective or transmit beamforming. This validates the effectiveness of joint beamforming in enhancing detection performance. 
\vspace{-7pt}
\section{Conclusion}
\vspace{-5pt}
This letter studied the active IRS-enabled target detection, with the objective of detecting the presence of a target by processing echo signals through the BS-IRS-target-IRS link. Utilizing the NP theorem, we designed an optimal detector by jointly leveraging the BS's transmit signal and the active IRS’s reflection noise. After that, we jointly optimized the transmit and reflective beamforming to maximize the detection probability. Simulation results demonstrated that signal amplification at the active IRS, the utilization of the IRS’s reflection noise for detector design, and the joint beamforming optimization can significantly increase detection probability.

\vspace{-7pt}
\appendix
\vspace{-5pt}
\subsection{Proof of Proposition~\ref{lm:P_D_under_P_F}}\label{sec:Proof_of_Lemma_1}
First, we consider the case when the target is absent. In this case, the detector $T(\tilde{\mathbf y})$ becomes 
$T(\tilde{\mathbf y};\mathcal{H}_0)= 
-  \sum_{t=1}^T\left|\frac{(1-k_1M_r)\alpha\mathbf e^T(\theta_0,N)\mathbf A\mathbf \Phi\mathbf G \mathbf x(t)}{\sigma\sqrt{k_1}}  \right|^2+ \sum_{t=1}^T\left|\frac{\sqrt{k_1}}{\sigma} \mathbf e^H(\theta_0,M_r)\mathbf n(t)+ \frac{(1-k_1M_r)}{\sigma\sqrt{k_1}} \alpha\mathbf e^T(\theta_0,N)\mathbf A\mathbf \Phi\mathbf G \mathbf x(t)\right|^2$
% \begin{equation}
% \begin{split}
% \quad T(\tilde{\mathbf y};\mathcal{H}_0)= &
% -  \sum_{t=1}^T\left|\frac{(1-k_1M_r)\alpha\mathbf e^T(\theta_0,N)\mathbf A\mathbf \Phi\mathbf G \mathbf x(t)}{\sigma\sqrt{k_1}}  \right|^2\\
% & + \sum_{t=1}^T\left|\frac{\sqrt{k_1}}{\sigma} \mathbf e^H(\theta_0,M_r)\mathbf n(t)\right.\\
% &\left.+ \frac{(1-k_1M_r)}{\sigma\sqrt{k_1}} \alpha\mathbf e^T(\theta_0,N)\mathbf A\mathbf \Phi\mathbf G \mathbf x(t)\right|^2,
% \end{split}
% \end{equation}
with $k_1 = \frac{|\alpha|^2\sum_{n=1}^N a_n^2}{\sigma^2/\sigma_z^2+|\alpha|^2M_r\sum_{n=1}^N a_n^2}$.
Due to the fact that $\mathbf e^H(\theta_0,M_r)\mathbf n(t)\sim \mathcal{CN}\left(\mathbf 0, M_r\sigma^2\right)$, by letting $\mathcal P(\theta_0) = \mathbf e^T(\theta_0,N)\mathbf A\mathbf \Phi\mathbf G \mathbf R_x \mathbf G^H \mathbf \Phi^H\mathbf A^H \mathbf e^*(\theta_0,N)$, we construct an equivalent detector of $T(\tilde{\mathbf y};\mathcal{H}_0)$ as 
$T'(\tilde{\mathbf y};\mathcal{H}_0)=\frac{T(\tilde{\mathbf y};\mathcal{H}_0) +  \frac{(1-k_1M_r)^2|\alpha|^2T}{\sigma^2k_1}\mathcal P(\theta_0)}{k_1M_r/2}$, which is the sum squares of independent and identically distribution (i.i.d.) Gaussian random vectors with mean values $\sqrt{\frac{2}{M_r}}\frac{(1-k_1M_r)}{\sigma{k_1}}\alpha \mathbf e^T(\theta_0,N)\mathbf A\mathbf \Phi\mathbf G \mathbf x(t), t\in\mathcal{T}$ and variance $1$, and thus follows non-central chi-squared distribution with PDF
\begin{equation}
p_1(x) \!=\! \begin{cases}
\frac{1}{2}\left(\frac{x}{\lambda}\right)^{\frac{\nu_1-2}{4}}\!\exp\left[-\frac{1}{2}\!(x\!+\!\lambda_1)\right] I_{\frac{\nu_1}{2}-1}(\sqrt{\lambda_1 x}), & x>0,\\
0, &x<0,
\end{cases}
\end{equation}
where the degrees of freedom are $\nu_1 = 2T$, the non-centrality parameter is $\lambda_1 = \frac{2\sigma^2T\mathcal P(\theta_0)}{(\sum_{n=1}^Na_n^2)^2M_r\sigma_d^4|\alpha|^2}$, and $I_r(u)$ is the modified Bessel function of the first kind and order $r$. Accordingly, the false alarm probability $P_\text{FA}$ given threshold $\delta'$ is given as $P_\text{FA} = \mathcal{Q}_{\chi^2_{2T}(\lambda_1)}\left(\frac{2}{k_1M_r}\delta' + \lambda_1\right)
$,
% \begin{equation}\label{eq:FA_probability}
% P_\text{FA} = \mathcal{Q}_{\chi^2_{2T}(\lambda_1)}\left(\frac{2}{k_1M_r}\delta' + \lambda_1\right),
% \end{equation}
where $\mathcal{Q}_{\chi^2_{2T}(\lambda_1)}(x) = \int_{x}^{\infty}p_1(x)dx,x>0$. 

Next, we consider the case when the target is present and derive the detection probability. In this case, the detector $T(\tilde{\mathbf y})$ becomes $T(\tilde{\mathbf y};\mathcal{H}_1)=- \sum_{t=1}^T \left|\frac{(1-k_1M_r)\alpha\mathbf e^T(\theta_0,N)\mathbf A\mathbf \Phi\mathbf G \mathbf x(t)}{\sigma\sqrt{k_1}} \right|^2+ \sum_{t=1}^T\left|\frac{\sqrt{k_1}}{\sigma} \mathbf e^H(\theta_0,M_r)\mathbf y(t)\frac{(1-k_1M_r)}{\sigma\sqrt{k_1}} \alpha\mathbf e^T(\theta_0,N)\mathbf A\mathbf \Phi\mathbf G \mathbf x(t)\right|^2$.
% \begin{equation}
% \begin{split}
% T(\tilde{\mathbf y};\mathcal{H}_1)=&- \sum_{t=1}^T \left|\frac{(1-k_1M_r)\alpha\mathbf e^T(\theta_0,N)\mathbf A\mathbf \Phi\mathbf G \mathbf x(t)}{\sigma\sqrt{k_1}} \right|^2\\
% &+ \sum_{t=1}^T\left|\frac{\sqrt{k_1}}{\sigma} \mathbf e^H(\theta_0,M_r)\mathbf y(t)\right.\\
% &+ \left.\frac{(1-k_1M_r)}{\sigma\sqrt{k_1}} \alpha\mathbf e^T(\theta_0,N)\mathbf A\mathbf \Phi\mathbf G \mathbf x(t)\right|^2.
% \end{split}
% \end{equation}
As $\mathbf e^H(\theta_0,M_r)\mathbf y(t)\sim \mathcal{CN}(\alpha M_r\mathbf e^T(\theta_0,M_r)\mathbf A\mathbf \Phi \mathbf G\mathbf x(t), M_r\sigma^2+|\alpha|^2M_r^2\sum_{n=1}^Na_n^2\sigma_z^2 )$, we construct an equivalent detector of $T(\mathbf y;\mathcal{H}_1)$ as 
$T'(\tilde{\mathbf y};\mathcal{H}_1)=\frac{T(\tilde{\mathbf y};\mathcal{H}_1) +  \frac{(1-k_1M_r)^2|\alpha|^2T}{\sigma^2k_1}\mathcal P(\theta_0)}{k_1\left(M_r+|\alpha|^2M_r^2\sum_{n=1}^Na_n^2\sigma_z^2/\sigma^2\right)/2}$, 
which follows non-central chi-squared distribution with PDF
\begin{equation}
p_2(x) \!=\! \begin{cases}
\frac{1}{2}\!\left(\frac{x}{\lambda}\right)^{\frac{\nu_2-2}{4}}\!\exp\left[-\frac{1}{2}(x\!+\!\lambda_2)\right] I_{\frac{\nu_2}{2}-1}(\sqrt{\lambda_2 x}), & x>0,\\
0, &x<0,
\end{cases}
\end{equation}
where the degrees of freedom are $\nu_2 = 2T$ and the non-centrality parameter is  $\lambda_2 = \lambda_1(1+|\alpha|^2\sum_{n=1}^Na_n^2M_r\sigma_z^2/\sigma^2)$. Accordingly, the detection probability $P_\text{D}$ given threshold $\delta'$ is given as $P_\text{D} = \mathcal{Q}_{\chi^2_{2T}(\lambda_2)}\Big(\frac{2}{k_1\left(M_r+|\alpha|^2M_r^2\sum_{n=1}^Na_n^2\sigma_z^2/\sigma^2\right)}\delta' + \lambda_1\left(1+|\alpha|^2\sum_{n=1}^Na_n^2M_r\sigma_z^2/\sigma^2\right)\Big),$
% \begin{equation}\label{eq:TD_probability}
% \begin{split}
% P_\text{D} =&~ \mathcal{Q}_{\chi^2_{2T}(\lambda_2)}\Bigg(\frac{2}{k_1\left(M_r+|\alpha|^2M_r^2\sum_{n=1}^Na_n^2\sigma_z^2/\sigma^2\right)}\delta'\Bigg.\\
% &\Bigg. + \lambda_1\left(1+|\alpha|^2\sum_{n=1}^Na_n^2M_r\sigma_z^2/\sigma^2\right)\Bigg),
% \end{split}
% \end{equation}
where $\mathcal{Q}_{\chi^2_{2T}(\lambda_2)}(x) = \int_{x}^{\infty}p_2(x)dx,x>0$. 
Finally, we obtain the detection probability under given false alarm probability in \eqref{eq:P_D_given_P_F}.
\vspace{-5pt}
\subsection{Proof of Proposition~\ref{le:P_D_approximation}}\label{Sec:proof_lemma_2}
\begin{figure*}[t]
\begin{equation}\label{eq:D_probability_approx}
\tilde{P}_\text{D} = \mathcal{Q}\Bigg(\frac{\frac{2}{k_1\left(M_r+|\alpha|^2M_r^2\sum_{n=1}^Na_n^2\sigma_z^2/\sigma^2\right)}\delta'+\frac{\lambda_1}{1+|\alpha|^2M_r\sum_{n=1}^Na_n^2\sigma_z^2/\sigma^2}-\lambda_1\left(1+|\alpha|^2\sum_{n=1}^Na_n^2M_r\sigma_z^2/\sigma^2\right)-2T}{\sqrt{4T+4\lambda_2}}\Bigg).\tag{9}
\end{equation}
\hrulefill
\vspace{-10pt}
\end{figure*}
When $T$ is sufficiently large, the central limit theorem implies that the distributions of $T'(\tilde{\mathbf y};\mathcal{H}_0)$ and $T'(\tilde{\mathbf y};\mathcal{H}_1)$ approach Gaussian\cite{steven1993fundamentals}, i.e., $T'(\tilde{\mathbf y};\mathcal{H}_0)\approx T''(\tilde{\mathbf y};\mathcal{H}_0)\sim \mathcal{N}(v_1+\lambda_1,2v_1+4\lambda_1)$ and $T'(\tilde{\mathbf y};\mathcal{H}_1)\approx T''(\tilde{\mathbf y};\mathcal{H}_1)\sim \mathcal{N}(v_2+\lambda_2,2v_2+4\lambda_2)$. Accordingly, the false alarm probability $P_\text{FA}$ and detection probability $P_\text{D}$ becomes $\tilde{P}_\text{FA} = \mathcal{Q}\left(\frac{\frac{2}{k_1M_r}\delta' -2T}{\sqrt{4T+4\lambda_1}}\right)$ and $\tilde{P}_\text{D}$ in \eqref{eq:D_probability_approx} as given at the top of this page. Based on these, the detection probability under a predetermined false alarm probability is given in \eqref{eq:P_D_approximation}.

\ifCLASSOPTIONcaptionsoff
  \newpage
\fi
\vspace{-5pt}
\bibliographystyle{IEEEtran}
\bibliography{IEEEabrv,mybibfile}

\end{document}